\documentclass[aps,prl,twocolumn]{revtex4}
\usepackage{graphicx}

\begin{document}
\title{Electron Spin Injection at a Schottky Contact}
\author{J. D. Albrecht and D. L. Smith}
\affiliation{Los Alamos National Laboratory, Los Alamos, New Mexico 87545}

\begin{abstract}
We investigate theoretically electrical spin injection at a Schottky contact
between a spin-polarized electrode and a non-magnetic semiconductor. 
Current and electron density spin-polarizations are discussed as
functions of barrier energy and semiconductor doping density. The effect of
a spin-dependent interface resistance that results from a tunneling region
at the contact/semiconductor interface is described. The model can serve
as a guide for designing spin-injection experiments with regard to the
interface properties and device structure.
\end{abstract}

\pacs{73.50.-h, 73.40.Qv, 73.30.+y}
\maketitle

%
%
%

Semiconductor device concepts that exploit the electron spin degree of
freedom require an electrical means of injecting spin-polarized currents
into a semiconductor. The two main experimental structures for meeting
this requirement use injection from a ferromagnetic metal or from a
spin-polarized semiconductor contact into a non-magnetic semiconductor. 
Such contacts are being studied both for their fundamental physics
properties as well as for a range of technological possibilities \cite
{reviewart}. Measurements of spin-polarized electron injection are often
made using a spin-LED configuration. In these experiments, electrons are
injected into an n-type semiconductor from a polarized contact and are
transported to a region in space, typically a quantum well, where they
recombine with nominally unpolarized holes transported from an adjacent
p-type doped region. The relative intensity of right- and left- circularly
polarized light emitted from the quantum well gives a measure of the
spin-polarization of the electron density in the recombination region. 
Recent measurements using injection from ferromagnetic contacts \cite
{ploogmetal,crowell,jonkermetal} and from spin polarized diluted magnetic
semiconductors contacts \cite{jonkerhet,molenhet} have been reported.

Theoretical discussion of spin injection has centered around a conductivity
mismatch between the contact and the semiconductor that can limit
polarization of the injected carriers. These considerations were presented
by Schmidt and coworkers \cite{schmidt}. Smith and Silver \cite{smithsilver}
subsequently included the possibility of a spin selective interface
resistance that results from tunneling and can improve spin injection.
Rashba formulated the problem in terms of an injection coefficient in which
currents dominated by tunneling at the interface can overcome the
limitations of a conductivity mismatch \cite{rashba}. These existing
theories treat the contact and semiconductor simply as uniform conductive
media and do not address critical issues of the real structures used in
experiments which typically consist of a Schottky contact with band bending
in a depletion region.

Here, we present a model of spin-polarized electron injection from a
reverse-biased Schottky contact. We analytically solve spin-dependent
continuity and drift-diffusion equations in the depletion region and examine
the influence of the interface and the depletion region on the
spin-polarized current and carrier densities in the semiconductor. We
include the possibility of a spin selective interface resistance that
results from tunneling processes at a ferromagnetic contact \cite{tunnel}. 
We emphasize the important distinction
between spin-polarization of the injected electron current and of the
electron density. Even if an injected current is highly polarized it can
result in small changes in the spin population of conduction electrons if
the electron gas into which injection occurs has a high density or the
magnitude of the injection current is small.

\begin{figure}
\rotatebox{0}{
\includegraphics{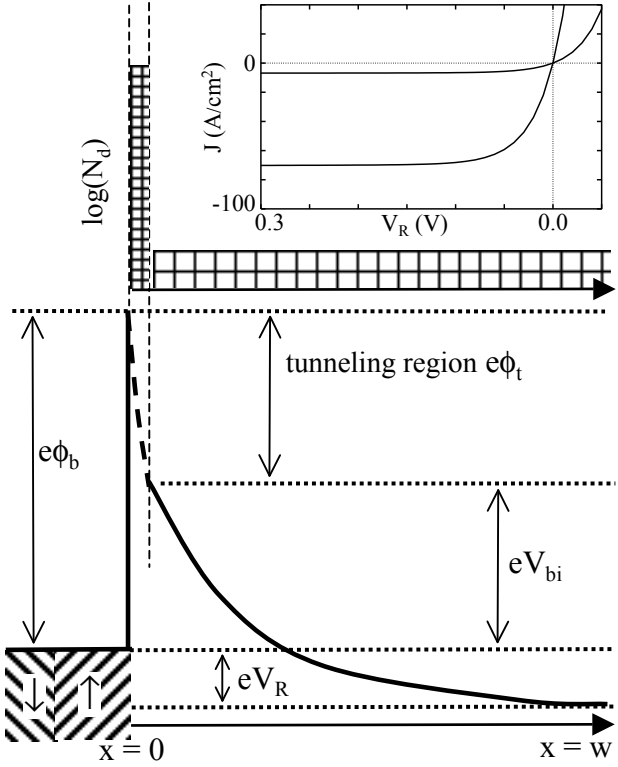}
}
\caption{\label{fig1}Energy diagram of a Schottky contact including the possibility of
a narrow tunneling region near the interface. The highly doped region near
the interface, through which electrons tunnel, is indicated by the dashed
portion of the conduction band profile. The corresponding doping profile
is shown above. Two calculated diode characteristics are inset for $V_{bi}$$=$0.2V
and $N_{d}$$=$$10^{16}$cm$^{-3}$ (smaller reverse saturation current) and $%
10^{17}$cm$^{-3}$.}
\end{figure}

An energy diagram for a Schottky barrier, which includes the possibility of
a narrow tunneling region near the interface, is shown in Fig.\ \ref{fig1}. A heavily
doped region near the interface, as illustrated by the doping profile in the
upper panel of Fig.\ \ref{fig1}, can be designed to form a sharp potential profile
through which electrons tunnel. The heavily doped region reduces the
effective Schottky energy barrier that determines the properties of the
depletion region \cite{sze}. The total barrier e$\phi _{b}$ is divided into
two parts, a tunneling region with barrier height e$\phi _{t}$ and an
effective Schottky barrier height $eV_{bi}$. The potential drop in the
depletion region consists of the effective Schottky barrier height plus the
applied reverse bias $eV_{R}$. Two parameters of the tunneling region, its
tunneling resistance and the magnitude of the reduction of the effective
Schottky barrier, can be separately controlled by the parameters of the
doping profile, for example the height and width of the heavily doped
region. The inset of Fig.\ \ref{fig1} shows calculated current-voltage
characteristics for two Schottky contacts with different bulk doping levels.
Spin injection experiments are typically performed in reverse bias in which
electrons are transported from the contact to the semiconductor.

The calculation decouples into a part for charge currents and densities and
a part for spin currents and densities. The calculation for charge currents
and densities is standard. We use a depletion approximation for the
electrostatics and the diffusion/thermionic emission model for the electron
current and density \cite{sze2}. We treat the spin current components using
drift-diffusion equations
\begin{equation}
j_{\eta }=\sigma _{\eta }\frac{\partial \left( \mu _{\eta }/e\right) }{%
\partial x}  \label{eq1}
\end{equation}
where $j_{\eta }$ is the current density, $\sigma _{\eta }$ is the
conductivity, and $\mu _{\eta }$ is the electrochemical potential for
electrons of spin type $\eta$$=$$\uparrow$$,$$\downarrow$. 
In the depletion region the conductivity
varies with the local electron concentration 
$n_{\eta }$$=$$\frac{1}{2}n_{i}\exp %
\left[ \left( e\phi +\mu _{\eta }\right) /kT\right] $. 
The contact and bulk 
semiconductor outside the depletion region are taken to be uniformly conducting
and the electrochemical potentials
relax to equilibrium in these extended regions according to $\partial ^{2}\mu _{-}/\partial
x^{2}$$=$$\mu _{-}/\Lambda ^{2}$\  where 
$\mu _{\uparrow }$$-$$\mu _{\downarrow }$$=$$\mu _{-} $ and $\Lambda $\ is 
the spin-diffusion length in the contact or semiconductor. Because of the large
electric field and rapidly varying electron density
in the depletion region, a spin diffusion equation is not
valid and we use spin-dependent continuity equations. Taking the difference
in the continuity equations for the two spin types gives, 
\begin{equation}
\frac{\partial \left( j_{\uparrow }-j_{\downarrow }\right) }{\partial x}=%
\frac{en_{i}}{\tau _{s}}e^{e\phi /kT}\Omega  \label{eq2}
\end{equation}
where $\tau _{s}$\ is the spin lifetime in the semiconductor, $n_{i}$\ is
the intrinsic carrier density, and $\Omega$$=e^{\mu _{\uparrow }/kT}$$-$$e^{\mu
_{\downarrow }/kT}$. 
The spin lifetime and spin diffusion length are related by 
$\Lambda ^{2}$$=$$(kT/e)$$\bar{\mu}$$\tau _{s}/2$.
The electron mobility is $\bar{\mu}$ and the $\frac{1}{2}$ appears because of
particle conservation.
Taking the difference in the drift-diffusion equations
for the two spin types gives 
\begin{equation}
j_{\uparrow }-j_{\downarrow }=\frac{\bar{\mu}n_{i}kT}{2}e^{e\phi /kT}\frac{%
\partial \Omega }{\partial x}  \label{eq4}.
\end{equation}
Given that the
electrostatic potential in the depletion region is quadratic, Eqs.\ \ref{eq2}
and \ref{eq4} can be combined to give an equation of the form \cite{limitcase} 
\begin{equation}
\frac{\partial ^{2}\Omega }{\partial x^{2}}+(-ax+b)\frac{\partial \Omega }{%
\partial x}-\frac{\Omega}{\Lambda ^{2}} =0  \label{eq5}
\end{equation}
where $a$ and $b$ are known constants that follow from the
electrostatic solution in the depletion region.
Eq.\ \ref{eq5} can be transformed to a confluent hypergeometric
equation by a change of variables and thus solved analytically in terms of
two matching coefficients \cite{trans}. These coefficients are determined by
matching to the solutions for $\mu
_{-} $ in the contact and in the charge-neutral region outside of the
depletion region. Once the matching coefficients are known, the spin
polarized currents and electron densities can be calculated.
A spin-dependent interface resistance is incorporated to describe
tunneling as in Ref.\ [\onlinecite{smithsilver}].

\begin{figure}
\rotatebox{0}{
\includegraphics{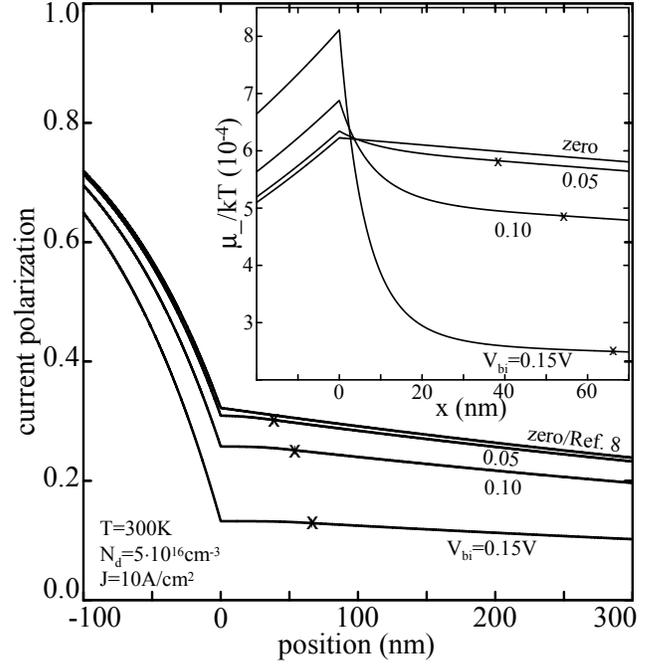}
}
\caption{\label{fig2}Current polarization as a function of position for various Schottky
barrier heights. The inset shows the difference in
electrochemical potentials near the interface.}
\end{figure}

The model can be applied both to metal / semiconductor contacts and to
heterojunction contacts with injection from a heavily doped, spin-polarized
semiconductor into a less heavily doped unpolarized semiconductor with a
higher energy conduction band \cite{awschalom}. We first consider parameters
appropriate to the heterostructure case. In Fig.\ \ref{fig2} we show the calculated
spin current polarization, $(j_{\uparrow }$$-$$j_{\downarrow })/(j_{\uparrow
}$$+$$j_{\downarrow })$, as a function of position for a series of structures
with different barrier heights (negligibly small, 0.05, 0.1 and 0.15 eV), an
injection current density of $10$A$\cdot $cm$^{-2}$, and a bulk doping of $%
5$$\cdot$$10^{16}$cm$^{-3}$. The zero of position is the interface and the contact
(semiconductor) at negative (positive) values of $x$. 
The symbol x on the curves indicates the edge of
the depletion region. Results from Ref.\ [\onlinecite{smithsilver}] for the
same parameters are also shown. The contact is taken to be 95\% spin
polarized and with a conductivity twice that of the collecting
semiconductor. It is assumed that the contact has a lower mobility but is
more heavily doped than the collecting semiconductor so that depletion
occurs in the collecting semiconductor. The interface resistance is zero. A
mobility of $\bar{\mu}$=5000cm$^{2}$/Vs, a spin diffusion length of 1$\mu $m for the
collecting semiconductor, and a spin diffusion length of 100nm in the contact
at T=300K are used throughout the paper. The top two curves, which are
indistinguishable, show the calculation for negligible barrier height and for the 
constant conductivity model of Ref.\ [\onlinecite{smithsilver}] which does not have a
depletion region. In the limit of small energy barrier we
recover the results of the constant conductivity model.
There is a strong decrease in spin injection with
increasing barrier height for fixed doping. The inset of Fig.\ \ref{fig2} shows
the difference in electrochemical potentials for spin-up and spin-down
electrons, $\mu _{-}$, as a function of position. 
As the barrier height increases there is a rapid
drop in the difference in electrochemical potentials for spin-up and
spin-down electrons across the depletion region. This rapid drop in $\mu _{-}
$ across the depletion region is the cause of the decreased spin injection
with increasing barrier height seen in the upper panel of Fig.\ \ref{fig2}. The drop
results because the depletion region has a low and rapidly varying electron
density.

\begin{figure}
\rotatebox{0}{
\includegraphics{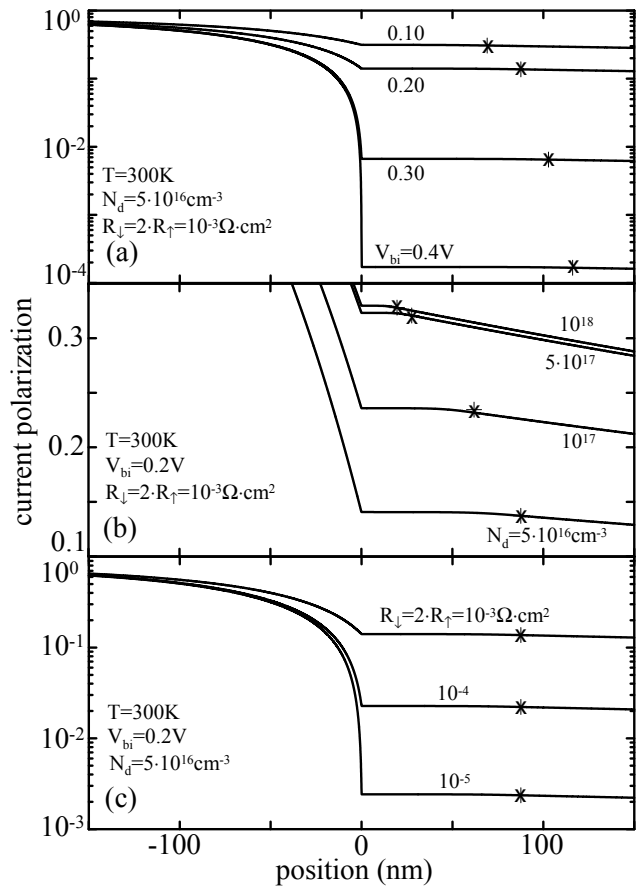}
}
\caption{\label{fig3}Current polarization as a function of position for (a) various
barrier height with fixed bulk doping, (b) for various bulk doping with
fixed barrier height, and (c) for various interface resistance values with
fixed doping and barrier height.}
\end{figure}

The heterostructure situation depicted in Fig.\ \ref{fig2} is somewhat idealized in
the sense that spin polarized n-type semiconductor injectors that do not
require high magnetic fields and low temperatures are still being sought.
However, it is feasible to grow ferromagnetic metals on semiconductors, for
example, epitaxial films of Fe on GaAs. In Fig.\ \ref{fig3} we show the calculated
current spin-polarization\ as a function of position from a metallic contact
(contact resistivity equal to 10$^{-5}$$\Omega$$\cdot$cm). We have, for
comparison purposes, computed all curves for 90\% of the reverse saturation
current density (which, of course, varies with barrier energy and bulk
doping). In Fig.\ \ref{fig3}a we show, for fixed bulk doping ($5$$\cdot$$10^{16}$cm$%
^{-3}$), current polarization curves corresponding to different effective
barrier heights (0.1, 0.2, 0.3, and 0.4eV) and a spin-selective resistance
at the interface of 10$^{-3}$$\Omega \cdot $cm$^{2}$ for spin-down current
and half this value for spin-up current \cite{intres}. A typical energy
barrier for Fe/GaAs is e$\phi _{b}$$\sim $0.7eV and we have assumed a barrier lowering
due to a heavily doped region near the interface. Fig.\ \ref{fig3}b shows an
analogous series of curves for a fixed energy barrier (0.2eV) and different
bulk doping densities ($5$$\cdot$$10^{16}$, $1$$\cdot$$10^{17}$, $5$$\cdot$$10^{17}
$, and $1$$\cdot$$10^{18}$cm$^{-3}$) with the same interface resistance. Fig.\
3c shows a series of curves in which the barrier height (0.2 eV) and bulk
doping ($5\cdot 10^{16}$cm$^{-3}$) are held fixed and the interface
resistance is varied ($10^{-3}$, $10^{-4}$, $10^{-5}$$\Omega \cdot $cm$^{2}$%
). From the results presented in Fig.\ \ref{fig3}, one sees that a depletion region
is highly undesirable for spin injection.  For efficient spin injection,
the effective barrier height should not exceed about 0.2eV.  Increasing the bulk
doping improves the current spin-polarization because it reduces the width of the 
depletion region.  It is also important to have a significant spin-dependent 
interface resistance.  Spin injection is sensitive to the doping profile. 
To maximize spin injection, a heavily doped region near the
interface should be used to reduce the effective energy barrier and form a 
spin-selective tunnel barrier to a ferromagnetic contact.

\begin{figure}
\rotatebox{0}{
\includegraphics{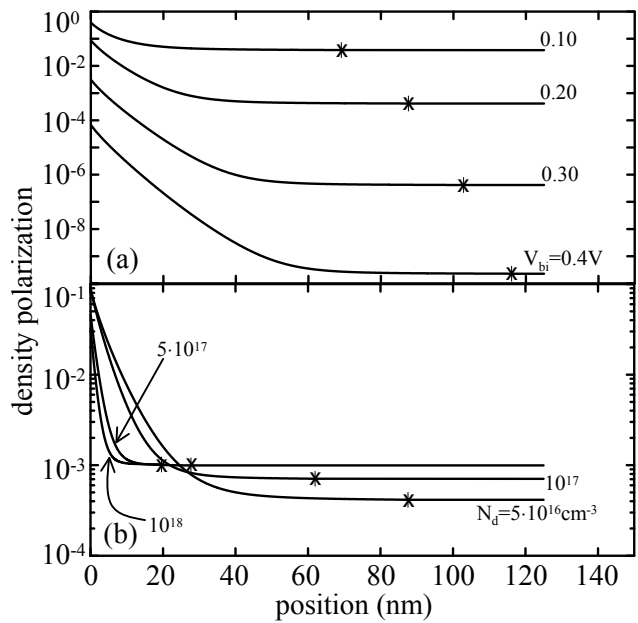}
}
\caption{\label{fig4}Electron density polarization as a function of position (a) for
various barrier height with fixed bulk doping and (b) for various bulk
doping with fixed barrier height. Parameters are as in Figs.\ \ref{fig3} (a) and (b).}
\end{figure}

Current polarization is not the only important issue for spin injection
experiments. A distinction should be made between the injected current
polarization and the polarization of the electron density. In the spin-LED
configuration, the observed degree of circularly polarized light is related
to the spin-polarization of the electron density at the region in space
where optical recombination occurs, typically in a quantum well. In Fig.\ \ref{fig4}a
we show, for the device parameters used in Fig.\ \ref{fig3}a, the electron density
spin-polarization $(n_{\uparrow }$$-$$n_{\downarrow })/(n_{\uparrow
}$$+$$n_{\downarrow })$\ as a function of position in the semiconductor for a
series of barrier heights at fixed doping. Only for the smallest effective
Schottky barrier is there a significant density polarization persisting tens
of nm into the semiconductor. This is the region of interest for
measurements of circularly polarized emission in the spin-LED configuration.
Fig.\ \ref{fig4}b shows the effect of varying bulk doping on the density
spin-polarization (parameters as for Fig.\ \ref{fig3}b). Even though both the
injection current density and the current polarization efficiency increase
with increased bulk doping concentrations, the higher density electron gas
becomes more difficult to polarize. There is a point of diminishing returns
on heavy bulk doping. To achieve significant electron density polarization
in the optical recombination region, the density there should be as low as
possible consistent with a small depletion region to ensure good spin
injection efficiency and large injection currents.

We have presented a model for electrical spin injection at a Schottky
contact between a spin-polarized electrode and a non-magnetic semiconductor.
We have found that a significant depletion region at a Schottky contact is
highly undesirable for spin injection. Design of the doping profile is very
important to maximize spin injection. A heavily doped region near the
interface can be used to form a sharp potential profile through which
electrons tunnel and which also reduces the effective Schottky energy
barrier that determines the properties of the depletion region. The doping
profile should be chosen so that the potential drop in the depletion region
is as small as possible, but the tunneling region must also have a
significant interface resistance (of order 10$^{-3}$$\Omega$$\cdot$cm$^{2}$).
Spin injection measurements using a spin-LED\ configuration are sensitive to
the electron density polarization in the optical recombination region. The
electron density in this recombination region should be as low as possible,
consistent with a small depletion region, so that it can be more easily spin
polarized.

Acknowledgment: We thank F. X. Bronold and I. Martin for helpful
discussions. This work was supported by the SPINs program of the Defense
Advanced Research Projects Agency.


\begin{thebibliography}{99}
\bibitem{limitcase}  Eq. \ref{eq5} can be shown to reduce to a spin diffusion equation
in the limit of uniform carrier density, small electric field, and $\mu _{-}$$\ll$$kT$.
\bibitem{trans}  Let $\Omega =(-ax+b)g(z)$ where $z=(-ax+b)^{2}/2a$. Then $%
g(z)$ satisfies a confluent hypergeometic equation (CHE) and has a solution
of the form $\alpha M+\beta U$ where $\alpha $ and $\beta $ are matching
coefficients and $M$ and $U$ are independent solutions to the CHE.
\bibitem{intres}  The interface resistance is small enough that it does not
significantly effect the reverse bias current-voltage characteristics.
\end{thebibliography}
\end{document}